\documentclass{aa}
\usepackage{graphicx}
\usepackage{rotfloat}
\usepackage{natbib}
\bibpunct{(}{)}{;}{a}{}{,}

\begin{document}
\title{A wide-field spectroscopic survey of the cluster of galaxies
  \object{Cl0024+1654}}  
\subtitle{II. A high--speed collision?}

\author{Oliver Czoske\inst{1,2}
  \and Ben Moore\inst{3}
  \and Jean-Paul Kneib\inst{1}
  \and Genevi\`eve Soucail\inst{1}
  }

\offprints{Oliver Czoske, \email{oczoske@ast.obs-mip.fr}}

\institute{
  Observatoire Midi-Pyr\'en\'ees, UMR5572, 14 Av. Edouard Belin, 
  31400 Toulouse, France
  \and Institute for Astronomy, University of Hawaii, 2680 Woodlawn
  Drive, Honolulu HI 96822, USA
  \and Department of Physics, Durham University, South Road, Durham,
  DH1~3LE, UK
  }

\date{Received yes, Accepted yes}

\authorrunning{O. Czoske et al.}
\titlerunning{Spectroscopic survey of Cl0024+1654: II.}

\abstract{ The mass distribution of the rich cluster of galaxies
  Cl0024+1654 has frequently been used to constrain the nature of dark
  matter yet a model consistent with all the observational data has
  been difficult to construct. In this paper we analyse the
  three-dimensional structure of this cluster using new spectroscopic
  information on $\sim\!300$ galaxies within a projected radius of
  $3\,h^{-1}\,\mathrm{Mpc}$.  These data reveal an unusual foreground
  component of galaxies separated from the main cluster by
  $3000\,\mathrm{km\,s}^{-1}$. We use numerical simulations to show
  that a high speed collision along the line of sight between
  Cl0024+1654 and a second cluster of slightly smaller mass can
  reproduce the observed peculiar redshift distribution. Such a
  collision would dramatically alter the internal mass distribution of
  the bound remnants, creating constant density cores from initially
  cuspy dark matter profiles and scattering galaxies to large
  projected radii, consistent with the observed distribution of
  galaxies in Cl0024+1654. The proposed scenario can reconcile the
  inferred mass profile from gravitational lensing with predictions
  from hierarchical structure formation models, while at the same time
  resolving the mass discrepancy that results from a comparison
  between lensing, velocity dispersion and X-ray studies.
  }

\maketitle
%
\section{Introduction}
\label{sec:introduction}

The nature of the ubiquitous dark matter remains one of the major
mysteries in cosmology. The traditional view is that the
dark matter is cold, i.~e.\ non-relativistic at the decoupling
epoch, and only subject to gravitational and weak
interactions. Structure formation with this type of dark matter has
been investigated extensively through numerical simulations over the
past 25 years. High-resolution simulations of individual cold dark
matter halos have shown that the mass distribution in CDM halos over a
wide range of total masses, from dwarf galaxy to cluster scales, should
follow a universal profile \citep[NFW]{Navarro1997}; in particular, the
mass distribution is found to have a central cusp with logarithmic
slope of $-1.5$ \citep{Moore1998, Ghigna2000}. Recent observational
results, however, have challenged this simple picture. 

The rich cluster of galaxies Cl0024+1654 at $z\!=\!0.395$ features a
spectacular quintuple gravitational arc system of radius
$106\,h^{-1}\,\mathrm{kpc}$ that has been used to reconstruct the
projected central mass distribution 
\citep{Kassiola1992, Smail1996, Tyson1998,
  Broadhurst2000}. Constructing a mass model for this cluster that is  
consistent with all the observations has proved very difficult,
however. The total projected mass enclosed within the arc radius is
fairly well constrained and yields a value of about
$1.6\times10^{14}\,h^{-1}\,M_{\sun}$. To reproduce this mass with an  
NFW type profile would require a cluster with characteristic velocity
dispersion larger than $2000\,\mathrm{km\,s}^{-1}$
\citep{Broadhurst2000, Shapiro-Iliev2000}. This is much higher than
observed in this or indeed in any other cluster of galaxies.  The mass
reconstruction by \citet{Tyson1998}, based on the detailed structure
of the gravitationally lensed background spiral galaxy, shows a flat
core in the projected central mass distribution with a core radius of 
$35\,h^{-1}\,\mathrm{kpc}$. This apparent contradiction to the
predictions of standard CDM simulations has prompted a number of
authors to investigate alternatives to the classic CDM scenario,
e.~g.\ warm dark matter, self-interacting dark matter,
etc. \citep{Spergel-Steinhardt2000, Hogan-Dalcanton2000, Moore2000}.

As more detailed observations of individual clusters of galaxies are
compiled, combinations of X-ray imaging and spectroscopy, velocity
dispersion measurements and lensing mass maps frequently reveal that
clusters that were thought to be simple relaxed objects are actually
more complex systems, frequently undergoing mergers or generally
showing signs of substructure and deviations from dynamical
equilibrium. Evidence for recent mergers or accretion is currently
accumulating through \textsc{Chandra} observations in the form of 
merger shocks 
\citep[e.~g.][]{Markevitch-Vikhlinin2001, Markevitch2001} or cold
fronts \citep{Markevitch2000, Vikhlinin2001, Mazzotta2001a}. Mergers
and substructure have also been invoked to explain the discrepancy between
mass estimates from different methods that is observed in many
clusters \citep{Miralda-Escude-Babul1995, Wu1998}.

Cl0024+1654 is an example of a cluster where the high mass inferred
from the strong lensing mass reconstructions is at variance with the
fairly low X-ray luminosity and temperature, that indicate a total
mass for the cluster which is a factor 2 to 3 smaller than the
lensing mass \citep{Soucail2000}. Both the galaxy distribution on the
sky and the X-ray morphology are regular and by themselves compatible
with a relaxed massive cluster, a notion which was further supported
by the high galaxy velocity dispersion of $\approx
1200\,\mathrm{km\,s}^{-1}$ found in the redshift surveys of
\citet{Dressler-Gunn1992} and \citet[hereafter D99]{Dressler1999}.

In \citet[Paper~I]{Czoske2001a} we presented a new catalogue of 650
measured redshifts (including those from D99) in a wide field around
Cl0024+1654. 300 galaxies in the catalogue have redshifts in the range
$0.37\!<\!z\!<\!0.42$, i.~e.\ are cluster members or lie in the
immediate neighbourhood of the cluster, and the redshift distribution
of these galaxies was found in Paper~I to be distinctly bimodal. In
the present paper we analyse the distribution of the galaxies in
redshift/real space and present a merger scenario which is able to
reconcile all the observations. Section \ref{sec:environment} recalls
the data used and describes the distribution of the redshifts globally
and as a function of position within the survey field, as well as the
distribution of spectral types based on equivalent widths of important
lines. In Section \ref{sec:scenario}, we develop the merger scenario
and use numerical simulations to show that it is possible to explain
the peculiar three-dimensional distribution of the galaxies in
Cl0024+1654. In Section \ref{sec:discussion} we show how the observed
structure is able to explain and resolve the discrepancy of mass
estimates of Cl0024+1654 derived from different observations. Finally,
in Section \ref{sec:conclusions} we summarize and conclude.

Throughout this paper we use a Hubble constant $H_0 \!=\!
100\,h^{-1}\,\mathrm{km\,s}^{-1}\,\mathrm{Mpc}^{-1}$ and assume an
Einstein-de~Sitter Universe with $\Omega_{\rm M}\!=\!1$ and
$\Omega_{\Lambda}\!=\!0$. At the redshift of Cl0024+1654,
$z\!=\!0.395$, $1\arcsec$ corresponds to
$3.195\,h^{-1}\,\mathrm{kpc}$.

\section{The cluster environment}
\label{sec:environment}

\subsection{The data}
\label{ssec:data}

In Paper~I we presented a new catalogue of photometric and
spectroscopic data for 679 objects in a wide field of
$21\!\times\!25\,\mathrm{arcmin}^2$ around Cl0024+1654. At the
redshift of the cluster, $z\!=\!0.395$, this corresponds to
$4\times4.8\,h^{-2}\,\mathrm{Mpc}^2$ (the true survey field is
delimited by an irregular polygon inscribed in this rectangle and
covers $16.8\,h^{-2}\,\mathrm{Mpc}^2$, see Fig.~\ref{fig:maps}, or
Fig.~7 in Paper~I). The catalogue lists equatorial position, redshift,
$V$ magnitude, $V\!-\!I$ colour, equivalent widths for
[\ion{O}{ii}]$\,\lambda 3727$, [\ion{O}{iii}]$\,\lambda5007$,
H$\alpha$ (where within the wavelength range), H$\beta$ and H$\delta$,
as well as the strength of the 4000\AA\ break. Here we are only
concerned with the 650 objects having measured redshifts, of which 581
were observed by us at CFHT and WHT, with the remaining 69 taken from
\citet{Dressler1999}. See Paper~I for more technical details on the
data used in this paper.

\subsection{Redshift distribution}
\label{ssec:z-distribution}

\begin{figure}
  \resizebox{\hsize}{!}{\includegraphics{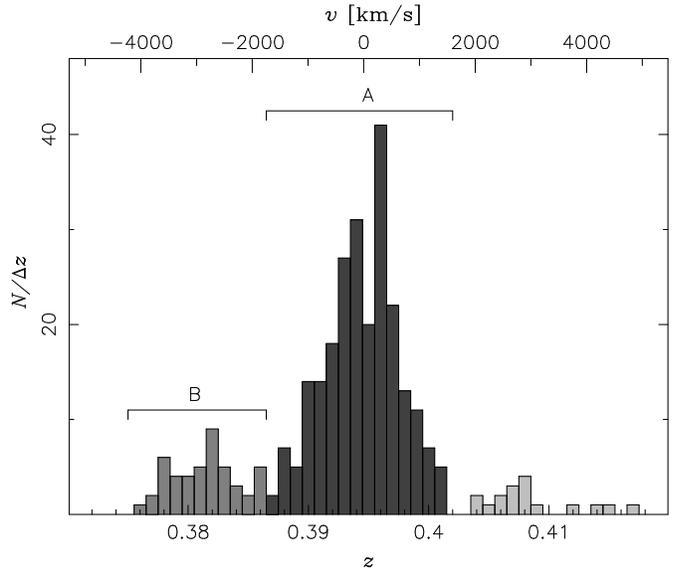}}
  \caption{Redshift histogram for 300 objects in the neighbourhood of
    Cl0024+1654 ($0.37\!<\!z\!<\!0.42$). The attribution of galaxies to
    components A (dark grey) and B (light grey) was done by inspection of
    Fig.~\ref{fig:scatterplot}. Whereas the lower and upper redshift 
    limits of components B and A respectively are fairly secure, the
    exact demarcation between the two components is arbitrary to some
    extent and was chosen at $\Delta v = -1500\,\mathrm{km\,s}^{-1}$
    with respect to the central redshift of component A, Eq.\
    (\ref{eq:mean_A}).}
  \label{fig:histogram}
\end{figure}

\begin{figure} 
  \resizebox{0.945\hsize}{!}{\includegraphics{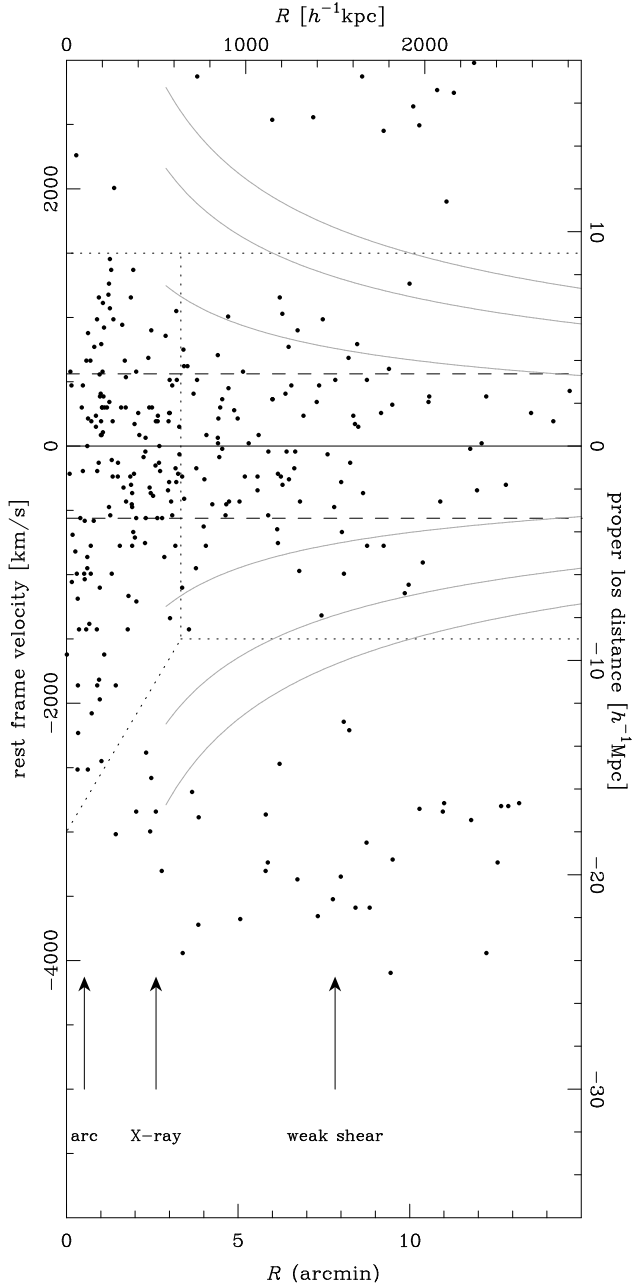}}
  \caption{Redshift $z$ plotted against angular distance $R$ from the
    projected cluster centre for the galaxies around the cluster
    redshift. The left axis expresses redshift as relative velocity
    with respect to the mean redshift of component A (Eq.\
    \ref{eq:mean_A}), the right axis as proper line-of-sight distance
    at the cluster redshift. Angular distance is converted to proper
    transverse distance on the top axis. The dashed horizontal lines
    indicate the velocity dispersion of component A (Eq.\
    \ref{eq:sigma_A}), the dotted lines denote limits for the samples
    used in Sect.\ \ref{ssec:z-distribution}. The solid curves mark
    the escape velocities for masses inside radius $R$ of 1, 3 and
    $5\times10^{14}\,M_{\sun}$. Arrows mark
    the position of the giant arc and the X-ray and weak shear
    detection limits.
    }
  \label{fig:scatterplot}
\end{figure}

\begin{figure} 
  \resizebox{\hsize}{!}{\includegraphics{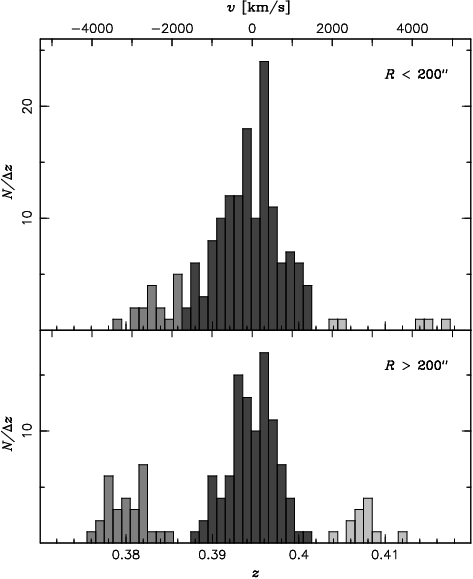}}
  \caption{Redshift histograms for objects inside (upper panel) and
    outside (bottom panel) a radius of $200\arcsec$
    (corresponding to $640\,h^{-1}\,\mathrm{kpc}$) from the projected
    cluster centre. In the central redshift distribution there is no
    correspondence to the distinct peak at $z\!=\!0.38$ visible in the
    external distribution. 
    }
  \label{fig:histo_slice}
\end{figure}

Fig.~\ref{fig:histogram} shows a histogram of the 300 redshifts that
lie in the range $0.37\!<\!z\!<\!0.42$. The
redshift distribution of the cluster galaxies is clearly bimodal,
showing two peaks at $z\!=\!0.381$ and $z\!=\!0.395$ respectively;
these peaks contain 283 galaxies. 

The larger peak at $z\!=\!0.395$ ($0.387\!<\!z\!<\!0.402$, hereafter
referred to as component A) contains 237 galaxies, is fairly regular
and resembles a Gaussian distribution as expected for a relaxed,
virialized cluster of galaxies. The smaller foreground peak at
$z\!=\!0.381$ ($0.374\!<\!z\!<\!0.387$, component B), by contrast,
seems too wide for the small number of 46 redshifts contained in
it. This impression is confirmed by Fig.~\ref{fig:scatterplot}, which
plots the redshift for each galaxy versus its projected distance from
the centre of Cl0024+1654\footnote{
  The coordinates in the catalogue are given relative to $\alpha_{2000}
  = 00^{\rm h}26^{\rm m}35\fs70$, $\delta_{2000} =
  17\degr09\arcmin43\farcs06$. Originally, this was the position of 
  galaxy 373, but a more accurate astrometric analysis shifted the
  reference point by $0\farcs79$ to the south-east of this galaxy.}.   
The distribution of the galaxies in the main peak is symmetrical with
respect to the central redshift line, whereas the distribution of the
foreground galaxies is roughly constant at a rest frame velocity of
$\sim\!-3000\,\mathrm{km}\,\mathrm{s}^{-1}$ at radii larger than
$3\arcmin$ ($600\,h^{-1}\,\mathrm{kpc}$), but turns off towards
smaller relative velocities to merge with the main distribution at
smaller projected distances. 

The 17 galaxies in the peak at $z\!=\!0.407$ are more widely dispersed
across the survey field and although we cannot rule out a connection
with Cl0024+1654, it seems more likely, in the light of the scenario
developed in Sect.\ \ref{sec:scenario}, that they are part of the 
surrounding field galaxy population.

It is remarkable that we can trace the main cluster as well as the
foreground structure out to the edge of the survey field (there
remains only one object in the catalogue outside the range depicted in
Fig.\ \ref{fig:scatterplot}, a background galaxy). The virial radius
for Cl0024+1654 should be around $1\,h^{-1}\,\mathrm{Mpc}$
\citep{Girardi-Mezzetti2001}, which means that we detect a coherent
structure out to three times the virial radius.

\subsubsection*{Redshift distribution inside $200\arcsec$}
\label{ssec:central_dist}

From Fig.\ \ref{fig:scatterplot} and Fig.\ \ref{fig:histo_slice}
(upper panel) it is obvious that the central region of Cl0024+1654 is
highly perturbed and it is impossible to separate components A and B
within $\sim\!200\arcsec$ ($640\,h^{-1}\,\mathrm{Mpc}$) from the
projected cluster centre. However, the redshift distribution for the
central galaxies is strongly skewed towards negative velocities:
except for two galaxies that are fairly isolated in redshift space,
the redshift distribution at positive velocities is effectively cut
off at $\sim\!  1500\,\mathrm{km\,s}^{-1}$, whereas the distribution
at negative velocities extends to beyond $\sim\!
-2500\,\mathrm{km\,s}^{-1}$, branching off into component B at $\sim\!
500\,h^{-1}\,\mathrm{kpc}$ from the projected cluster centre.
Quantitatively the skewness of the distribution of 161 galaxies within
$200\arcsec$ ($0.37\!<\!z\!<\!0.42$) is 0.54, where we use the
definition given by \citet{Press1992}:
\begin{equation}
  \label{eq:skewdef}
  \mathrm{Skew}(x_1\dots x_N) = \frac{1}{N} \sum_{j=1}^{N}
  \left[\frac{x_j-\overline{x}}{\sigma}\right]^3\quad.
\end{equation}
The probability that a sample of size $N\!=\!161$ drawn from a normal
distribution shows a skewness larger than that measured in Cl0024+1654
is about 0.3\%. The Shapiro-Wilk normality test
\citep{Shapiro-Wilk1965} rejects the hypothesis that these data points
are drawn from a Gaussian parent population at $\gg99$\% confidence.
The skew in the central redshift distribution is already apparent in
the histogram given by \citet{Dressler1999} (see also Section
\ref{sec:discussion}).

\subsubsection*{Redshift distribution outside $200\arcsec$}
\label{ssec:periph_dist}

In the distribution of the galaxies outside $200\arcsec$ (Fig.\ 
\ref{fig:scatterplot} and Fig.\ \ref{fig:histo_slice}, bottom panel)
the two modes are clearly separated. The mean redshift of the $71$
galaxies belonging to component A at projected distances
$200\arcsec\!<\!R\!<\!500\arcsec$ is
\begin{equation}
  \label{eq:mean_A}
  \overline{z_{\rm A}\rule[0mm]{0mm}{1.5ex}} = 0.3946\pm0.0007\quad. 
\end{equation}
In Fig.\ \ref{fig:slidavg} we show the two velocity dispersion
profiles for galaxies in component A with either positive or negative
velocities with respect to the mean redshift $\overline{z_{\rm A}}$,
calculated in a sliding bin containing 30 galaxies. Outside $3\arcmin$
($600\,h^{-1}\,\mathrm{kpc}$) both profiles are flat at the same level
of $\sigma \approx 600\,\mathrm{km\,s}^{-1}$.  The profile for
galaxies with negative velocities (i.~e.\ those moving towards us,
Fig.\ \ref{fig:slidavg}) rises all the way to about $45\arcsec$
($140\,h^{-1}\,\mathrm{kpc}$), reaching a velocity dispersion of
nearly $900\,\mathrm{km\,s}^{-1}$ in the innermost bin. The dispersion
profile of the galaxies with positive velocities, by contrast, drops
back to a value of about $600\,\mathrm{km\,s}^{-1}$ after having
reached a maximum value of about $800\,\mathrm{km\,s}^{-1}$ at
$2\arcmin$ from the centre. Although the galaxy velocities used in
Fig.\ \ref{fig:slidavg} have been restricted to
$|v|<1500\,\mathrm{km\,s}^{-1}$, the different behaviour of the
profiles for galaxies with positive and negative velocities
respectively reflects the skew in the central redshift distribution
and indicates the possible presence of a bulk motion component towards
us in the central galaxy distribution.

The exact value of the velocity dispersion of the $71$ galaxies within
the same distance range as used for Eq.\ (\ref{eq:mean_A}) and
velocity $|v|<1500\,\mathrm{km\,s}^{-1}$ relative to the central
redshift $\overline{z_{\rm A}}$ is
\begin{equation}
  \label{eq:sigma_A}
  \sigma_{\rm A} = 561_{-83}^{+95}\,\mathrm{km\,s}^{-1}\quad.
\end{equation}
Both $\overline{z_{\rm A}}$ and $\sigma_{\rm A}$ were computed using
the biweight estimator \citep{Beers1990}, the errors (95\% confidence
level) were estimated by bootstrap resampling.

The skewness of the distribution of these 71 redshifts is $-0.12$,
which should be compared to the standard deviation of skewness for
samples of this size drawn from a normal distribution, $0.27$. The
Shapiro-Wilk test does not provide evidence for deviations from a
Gaussian distribution.

\begin{figure*}
  \resizebox{\hsize}{!}{\includegraphics{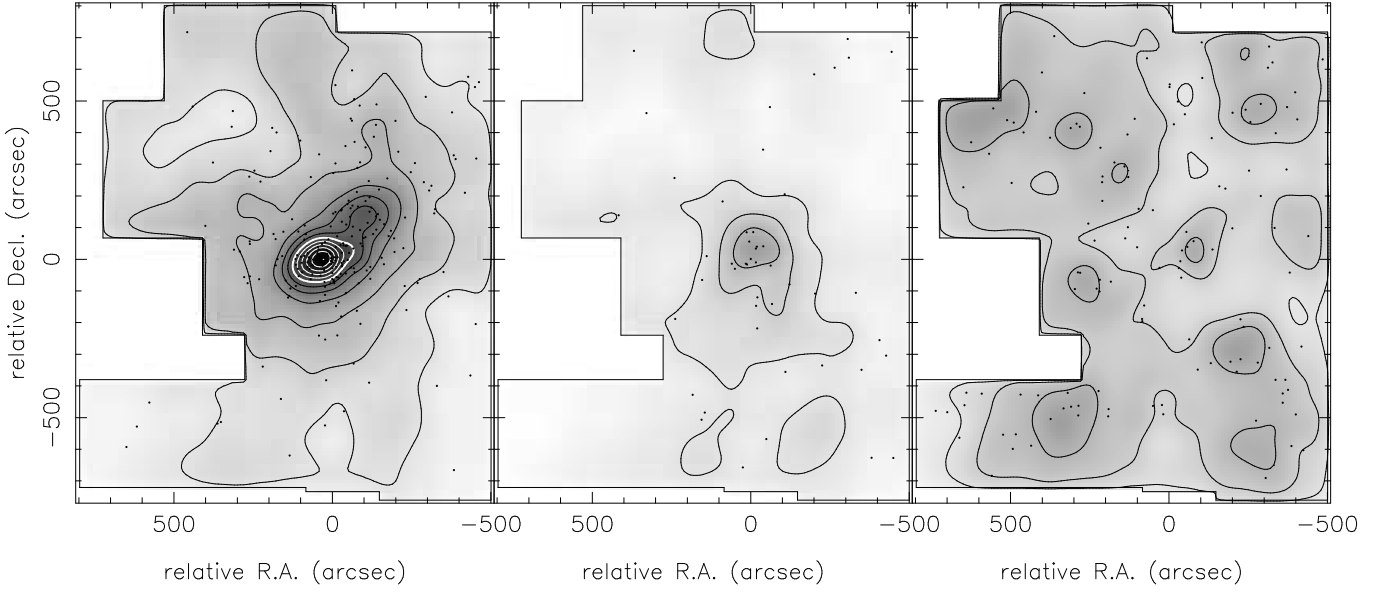}}
  \caption{Galaxy number density maps for (a) component A
    ($0.388 \!<\! z \!<\! 0.402$, 237 galaxies), (b) component B
    ($0.37 \!<\! z \!<\! 0.388$, 46 galaxies) and (c) the ``field''
    ($0.1 \!<\! z \!<\! 0.35$, 141 galaxies), estimated with the
    generalized nearest neighbour method \citep{Silverman1986} with 10 
    neighbours for map 
    (a) and 5 neighbours for maps (b) and (c). The maps are divided by
    the completeness map (see Paper I) and smoothed with a Gaussian
    with $\sigma\!=\!30\arcsec$. The 
    grey scales are the same for all three maps. The lowest density
    contour lines are at 2\% and 5\%, normalized to the maximum
    density in the cluster component A (left panel). The remaining
    contours are spaced in  steps of 10\%. For densities below 50\%
    the contours are drawn in black, for higher densities in
    white. The 50\% contour is marked by a bold white line.   
    }  
  \label{fig:maps}
\end{figure*}

\begin{figure}
  \resizebox{\hsize}{!}{\includegraphics{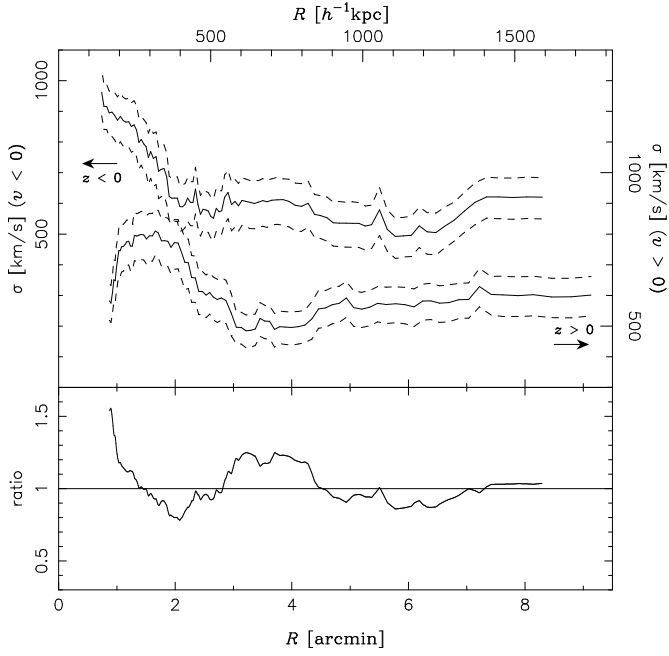}}
    \caption{Velocity dispersion profiles for galaxies with negative and
    positive velocity with respect to the mean redshift of cluster
    component A, which was held fixed at $z_{\rm A} = 0.3946$. The
    profiles were computed using sliding averages in bins containing
    30 galaxies. The error bands give $1\sigma$ errors determined from
    bootstrap resampling (10000 realisations). The curve for positive
    velocities was shifted by $300\,\mathrm{km\,s}^{-1}$ for clarity,
    with the corresponding velocity values marked on the right-hand
    axis. The bottom panel shows the ratio of the dispersions for
    galaxies with negative and positive velocities. 
    }
  \label{fig:slidavg}
\end{figure}

The spatial positions of the galaxies in the foreground component B
show no preference for any direction (Fig.\ \ref{fig:maps}, middle
panel). The formal velocity dispersion for component B in the same
radial distance range as used in Eq.\ (\ref{eq:sigma_A}) is
$\sigma_{\rm B}\!=\!554^{+175}_{-304}\,\mathrm{km\,s}^{-1}$ based on
15 redshifts. Given that the number of galaxies in component A is
almost five times as large as the number in component B it is hardly
conceivable that both components should have the same velocity
dispersion -- their mass-to-light ratios would be extremely different.
Component B is certainly not a virialized group or cluster.

From the distribution of the external galaxies alone one might still
suspect the presence of a loose aggregate of galaxies physically
disconnected from Cl0024+1654, but well-aligned with the line of
sight; however, in the centre there are no galaxies which would
correspond to the $z\!=\!0.38$ peak visible in the external
distribution (Fig.\ \ref{fig:histo_slice}). If the foreground
component B were an independent loose system, it would show a
deficiency of galaxies right in front of the centre of Cl0024+1654,
which seems unlikely in the absence of interaction between the
systems. This leads us to suspect that the two components are in fact
physically connected and that the blue tail of the skewed central
distribution originates from the same physical system as the
foreground component B.

Fig.\ \ref{fig:scatterplot} shows the theoretical escape velocities for
masses of 1, 3 and $5\times10^{14}\,M_{\sun}$ enclosed in a sphere of
radius $R$. Although these lines are merely indicative due to the
assumption of spherical symmetry and the unknown tangential velocity
components, it is safe to say that the galaxies in component B
outside, say, $600\,h^{-1}\,\mathrm{kpc}$ are not bound to the main
cluster component, unlike the negative tail of the skewed central
galaxy distribution.

The galaxy density map of the main component A (Fig.\ \ref{fig:maps})
shows an extension of the galaxy distribution towards the northwest to
a distance of $100\arcsec - 150\arcsec$. This extension is visible in
images of Cl0024+1654 and we show here that the galaxies contained in
it lie at the redshift of the main cluster. Cl0024+1654 might
therefore well be comprised of three distinct subsystems: the main
cluster, component A; a foreground cluster or group, component B,
which is interacting with the main cluster; and a third group or small
cluster, which is falling onto the main cluster in the plane of the
sky.

\subsection{Distribution of spectral types}
\label{ssec:equwidths}

\begin{table*}[htbp]
  \caption{Distribution of spectral types: classification by
    [\ion{O}{ii}] vs. H$\delta$ (the definition of the spectral types
    is taken from \citet{Balogh1999} and given in column 2). The
    fraction of galaxies of a given 
    spectral type is listed as the percentage of the total number of
    galaxies in the subsample (the total number is given in
    parentheses). See text and Fig.\ \ref{fig:scatterplot} for the
    definitions of the samples. The numbers are the percentages $p$ of
    galaxies belonging to each class in each sample. The errors are
    the standard deviations $\sqrt{Np(1-p)}$ of the normal
    distribution asymptotic to the binomial distribution with sample
    size $N$ and probability $p$ for large $N$.} 
  \begin{tabular}[h]{llcr@{$\pm$}lr@{$\pm$}lr@{$\pm$}lr@{$\pm$}lcrr}
    \hline\\[-2ex]
    & 
    &
    & \multicolumn{8}{c}{This work} 
    & 
    & \multicolumn{2}{c}{\citet{Balogh1999}} \\[0.5ex]
    \cline{4-11}  \cline{13-14}\\[-2ex]
    Spectral Type 
    & Definition
    & 
    & \multicolumn{2}{c}{Centre (183)} 
    & \multicolumn{2}{c}{A (64)}
    & \multicolumn{2}{c}{B (35)} 
    & \multicolumn{2}{c}{Field (275)} 
    & 
    & \multicolumn{1}{r}{Cluster} 
    & \multicolumn{1}{r}{Field} \\[0.5ex]
    \hline\\[-2ex]
    K+A$^\mathrm{a}$     
    & $\ion{H}{\delta}\!<\!-5\,$\AA, $[\ion{O}{ii}]\!<\!5\,$\AA 
    &  &  2 & 1 &  3 & 2 &  \multicolumn{2}{c}{0} &  4 & 1 & &  4  &  2 \\
    A+em$^\mathrm{b}$    
    & $\ion{H}{\delta}\!<\!-5\,$\AA, $[\ion{O}{ii}]\!>\!5\,$\AA
    &  & 3 & 1 &  6 & 3 & 11 & 5 &  3 & 1 & &  4 & 10 \\
    SF/SSB$^\mathrm{c}$  
    & $\ion{H}{\delta}\!>\!-5\,$\AA, $[\ion{O}{ii}]\!>\!5\,$\AA
    & & 23 & 3 & 42 & 6 & 57 & 8 & 53 & 3 & & 24 & 47 \\
    passive$^\mathrm{d}$
    & $\ion{H}{\delta}\!>\!-5\,$\AA, $[\ion{O}{ii}]\!<\!5\,$\AA
    & & 72 & 3 & 48 & 6 & 31 & 7 & 40 & 3 & & 67  & 41 \\[0.5ex]
    \hline
  \end{tabular}
  \\
  $^\mathrm{a}$spectrum dominated by K and A stars;
  $^\mathrm{b}$spectrum dominated by A stars, with emission lines;
  $^\mathrm{c}$star-forming/short starburst; $^\mathrm{d}$red,
  passively evolving galaxies
  \label{tab:oii-hd}
\end{table*}

\begin{table*}[htbp]
  \caption{Distribution of spectral types: classification  by
    H$\delta$ and the 4000\AA\ break. The structure of the table is the
    same as for Table \ref{tab:oii-hd}.}
  \begin{tabular}[h]{llcr@{$\pm$}lr@{$\pm$}lr@{$\pm$}lr@{$\pm$}lcrr}
    \hline\\[-2ex]
    & 
    &
    & \multicolumn{8}{c}{This work} 
    & 
    & \multicolumn{2}{c}{\citet{Balogh1999}} \\[0.5ex]
    \cline{4-11}  \cline{13-14}\\[-2ex]
    Spectral Type 
    & Definition 
    &
    & \multicolumn{2}{c}{Centre (148)} 
    & \multicolumn{2}{c}{A (60)}
    & \multicolumn{2}{c}{B (33)} 
    & \multicolumn{2}{c}{Field (225)} 
    & 
    & \multicolumn{1}{r}{Cluster} 
    & \multicolumn{1}{r}{Field} \\[0.5ex]
    \hline\\[-2ex]
    bHDS$^\mathrm{a}$    
    & $\ion{H}{\delta}\!<\!-5\,$\AA, $\mathrm{br} \!<\! 1.5$
    &  & 3 & 1 &  3 & 2 &  9 & 5 &  7 & 1 & &  5 &  7  \\
    rHDS$^\mathrm{b}$    
    & $\ion{H}{\delta}\!<\!-3\,$\AA, $\mathrm{br} \!>\! 1.5$
    &  & 8 & 2 & 17 & 5 &  9 & 5 &  5 & 1 & &  10 & 11  \\
    SF/SSB$^\mathrm{c}$  
    & $\ion{H}{\delta}\!>\!-5\,$\AA, $\mathrm{br} \!<\! 1.5$
    & & 34 & 3 & 50 & 6 & 67 & 8 & 56 & 3 & & 12 & 33 \\
    passive$^\mathrm{d}$ 
    & $\ion{H}{\delta}\!>\!-3\,$\AA, $\mathrm{br} \!>\! 1.5$
    & & 55 & 4 & 30 & 6 & 15 & 6 & 32 & 3 & & 74 & 45 \\[0.5ex]
    \hline
  \end{tabular}
  \\
  $^\mathrm{a}$blue \ion{H}{$\delta$} strong; $^\mathrm{b}$red
  \ion{H}{$\delta$} strong; $^\mathrm{c}$star-forming/short starburst;
  $^\mathrm{d}$red passively evolving galaxies
  \label{tab:br-hd}
\end{table*}

In Tables \ref{tab:oii-hd} and \ref{tab:br-hd} we classify the
galaxies into spectral types, defined as in \citet{Balogh1999}, using
our measured values for the equivalent widths of [\ion{O}{ii}],
\ion{H}{$\delta$} and the 4000\AA\ break (see Paper~I for details on
how these values are defined and measured).  

We consider four samples of galaxies according to redshift and
projected spatial position; the sample boundaries for samples
``Centre'', ``A'' and ``B'' are marked by the dotted lines in Fig.\
\ref{fig:scatterplot}. The boundary between ``Centre'' and ``B'' is to
some extent arbitrary; replacing the slanted line by a simple velocity
cut at $v\!=\!-1500\,\mathrm{km\,s}^{-1}$ does not significantly
change the numbers in Tables \ref{tab:oii-hd} and \ref{tab:br-hd}, 
however. ``Field'' includes all the galaxies in the sample with
redshifts $0 \!<\! z \!<\! 0.55$ without the cluster galaxies,  
$0.372 \!<\! z \!<\! 0.402$ (the upper limit here is the same as for
the field sample in \cite{Balogh1999}). 

Both the classifications by [\ion{O}{ii}] vs.\ \ion{H}{$\delta$} and
by \ion{H}{$\delta$} vs.\ the 4000~\AA\ break show an excess of blue
star-forming galaxies (star-forming/short starburst, SF/SSB, and
emission-line galaxies, A+em) in component ``B'' as compared to the
cluster centre and possibly even over the field. The distribution of
spectral types in component ``A'' is between a typical cluster and
field populations. This seems to provide support for the hypothesis
that component B is a loose galaxy overdensity with spectral
characteristics typical for field galaxies. Alternatively it is
possible that the star-formation activity in the galaxies belonging to
component B has recently been (re)activated and enhanced through
interaction with a denser environment.

We include in Tables \ref{tab:oii-hd} and \ref{tab:br-hd}
corresponding numbers from \citet{Balogh1999}. These numbers are not
strictly comparable, because the selection criteria used for the
creation of our catalogue are quite different from those used by
\citet{Balogh1999}\footnote{\citet{Balogh1999} use data from the CNOC1
  cluster redshift survey, which has limiting Gunn $r$ magnitudes
  between $20.5$ and $22.0$, depending on the redshift of the observed
  cluster \citep{Yee1996a}.}; 
this explains the discrepancies in particular in 
the \ion{H}{$\delta$} vs.\ the 4000~\AA\ break classification,
although the general trends are the same. The agreement in the
[\ion{O}{ii}] vs.\ \ion{H}{$\delta$} classification on the other hand
is very good, even quantitatively. 

\section{A high speed collision?}
\label{sec:scenario}

\begin{figure*}
  \caption{The initial and final particle configurations used to
    simulate the two colliding clusters. The grey scale indicates the
    local density of dark matter within the box of length 12 Mpc.}
  \label{fig:collide}
\end{figure*}

In general, a redshift difference between well-separated clusters at
cosmological distances is a combination of a cosmological redshift
difference due to the proper distance between the clusters along the
line of sight and a Doppler shift due to the relative velocities of
the two clusters. In general, these two effects cannot be disentangled
uniquely, and there might be several ways to reproduce a given
redshift distribution. In the following we develop a scenario that can
reproduce the observed redshift distribution in Cl0024+1654, in
particular the skewed central distribution, the presence of two
well-separated modes away from the projected cluster centre extending
to at least three times the virial radius.

Assume that a group or cluster has undergone a near radial collision
with Cl0024+1654 and we are now viewing the merging system along the
direction of the impact. In fact, several characteristic parameters of
a collision scenario are fairly well constrained by the observed
redshift distribution. The relative velocity of the two subclusters is
given by the redshift difference of components A and B and is thus of
order $3000\,\mathrm{km\,s}^{-1}$. Those parts of the smaller cluster
B that pass through the centre of A are more strongly decelerated in
the direction of the collision than its outer parts due to stronger
dynamical friction and they therefore show a smaller redshift
difference. A tidal gravitational shock during the crossing scatters
the outer galaxies of both clusters to large projected distance, thus
increasing the extent of the cluster halos beyond their nominal virial
radius. If the transverse velocity imparted on a galaxy that is now
found at a radial distance of $3h^{-1}\,\mathrm{Mpc}$ is
$1000\,\mathrm{km\,s}^{-1}$, then the time since core crossing is
about $3\,\mathrm{Gyrs}$, and the separation between the cores of the
two clusters is roughly $5\,\mathrm{Mpc}$.

The main unknown in this scenario is the mass ratio between the
subclusters. There are two possible versions of this collision: 
\begin{itemize}
\item[(i)] A small group has passed through the core of Cl0024+1654
  and has been completely disrupted and scattered by the cluster's
  tidal force. In this case the main body of Cl0024+1654 is
  unperturbed and we do not observe a concentration of foreground
  galaxies belonging to the original group because the system has been
  completely unbound and is scattered to large distances perpendicular
  to the main cluster.   

\item[(ii)] A massive cluster of approximately 50\% of the mass of
  Cl0024+1654 collided and passed through the core of the main cluster
  roughly $3\,\mathrm{Gyr}$ ago. This collision is insufficient to
  completely disrupt the impacting cluster but its outer galaxies are
  scattered to large projected distances.  The remaining bound core of
  the colliding cluster undergoes sufficient dynamical friction to
  turn its orbit around and in redshift space it appears to lie at a
  similar distance to Cl0024+1654.  
\end{itemize}
Both of these scenarios can reproduce the main features of the
observed redshift distribution. However, the massive collision
scenario is more interesting because of the possibility of resolving
both the mass discrepancy problem and the conflict between the central
density structure of Cl0024+1654 and predictions of hierarchical clustering
models. 

The mass reconstructions of Cl0024+1654 \citep{Tyson1998, Broadhurst2000}
using the positions of multiple images of a gravitationally lensed
background galaxy show a projected mass of 
$1.3\times10^{14}\,h^{-1}\,M_\odot$ within
$106\,h^{-1}\,\mathrm{kpc}$ and a central surface mass density of
$7900\,h\,M_\odot\,\mathrm{pc}^{-2}$. For a singular isothermal halo
extending to 
$r_{200}/\mathrm{kpc}\approx\sqrt{2}\,\sigma_\mathrm{1d}/(\mathrm{km\,s}^{-1})$  
this implies a characteristic velocity dispersion larger than
$1500\,\mathrm{km\,s}^{-1}$ in order to explain the projected mass.

As \citet{Shapiro-Iliev2000} point out, the situation is worse in the
context of hierarchical clustering models. Dark matter halos in cold
dark matter (CDM) type models are shallower than isothermal in their
centres. In order to reproduce the observed projected mass a CDM halo
with velocity dispersion of $2200\,\mathrm{km\,s}^{-1}$ is
required. This is inconsistent with the observations presented in this
paper.

A second problem for hierarchical models is that \citet{Tyson1998}
infer a very shallow central density profile for this cluster, much
flatter than the cuspy density profiles found for clusters in CDM type
models \citep{Ghigna2000}. The central core of Cl0024+1654 has
frequently been used to constrain the nature of dark matter and to
argue for alternative candidates to CDM \citep{Spergel-Steinhardt2000,
  Hogan-Dalcanton2000, Moore2000}.

A high speed encounter between two similar mass clusters can explain
all of the observations presented in Sect.\ \ref{sec:environment} and
reconcile the discrepancy between mass estimates derived for
Cl0024+1654. We now explore this scenario using high resolution
numerical simulations of colliding dark matter halos to study the
evolution of the mass distribution.  We construct two equilibrium CDM
halos with virial masses $9.5\times 10^{14}\,M_\odot$ and $5.0\times
10^{14}M_\odot$ with peak circular velocities
$v_\mathrm{peak}\!=\!1600$ and $1300\,\mathrm{km\,s}^{-1}$ and
concentrations $c\!=\!5$ and $c\!=\!7$ respectively.  Their initial
separation is $3\,\mathrm{Mpc}$ and relative velocity is
$-3000\,\mathrm{km\,s}^{-1}$.  The particle mass is set to
$5\times10^9\,M_\odot$ and we use an equivalent Plummer force
softening of $5\,\mathrm{kpc}$.    

After the collision we find that the outer regions of the smaller
cluster have become unbound and are streaming radially away from the impact
location.  The impulse velocity perpendicular to the encounter is of
the order $1000\,\mathrm{km\,s}^{-1}$.  Snapshots of the initial and
final times are shown in Fig.\ \ref{fig:collide}. The peak
circular velocities of the bound components have fallen to $1430$ and
$1140\,\mathrm{km\,s}^{-1}$ whilst the central 1D velocity dispersions
have reduced to $930$ and $710\,\mathrm{km\,s}^{-1}$
from initial values of $1033$ and $883\,\mathrm{km\,s}^{-1}$ respectively.
At the final time we find a total mass within a cylinder of radius
$106\,\mathrm{kpc}$  of $8.0\times 10^{13}M_\odot$.

\begin{figure}
  \resizebox{\hsize}{!}{\includegraphics{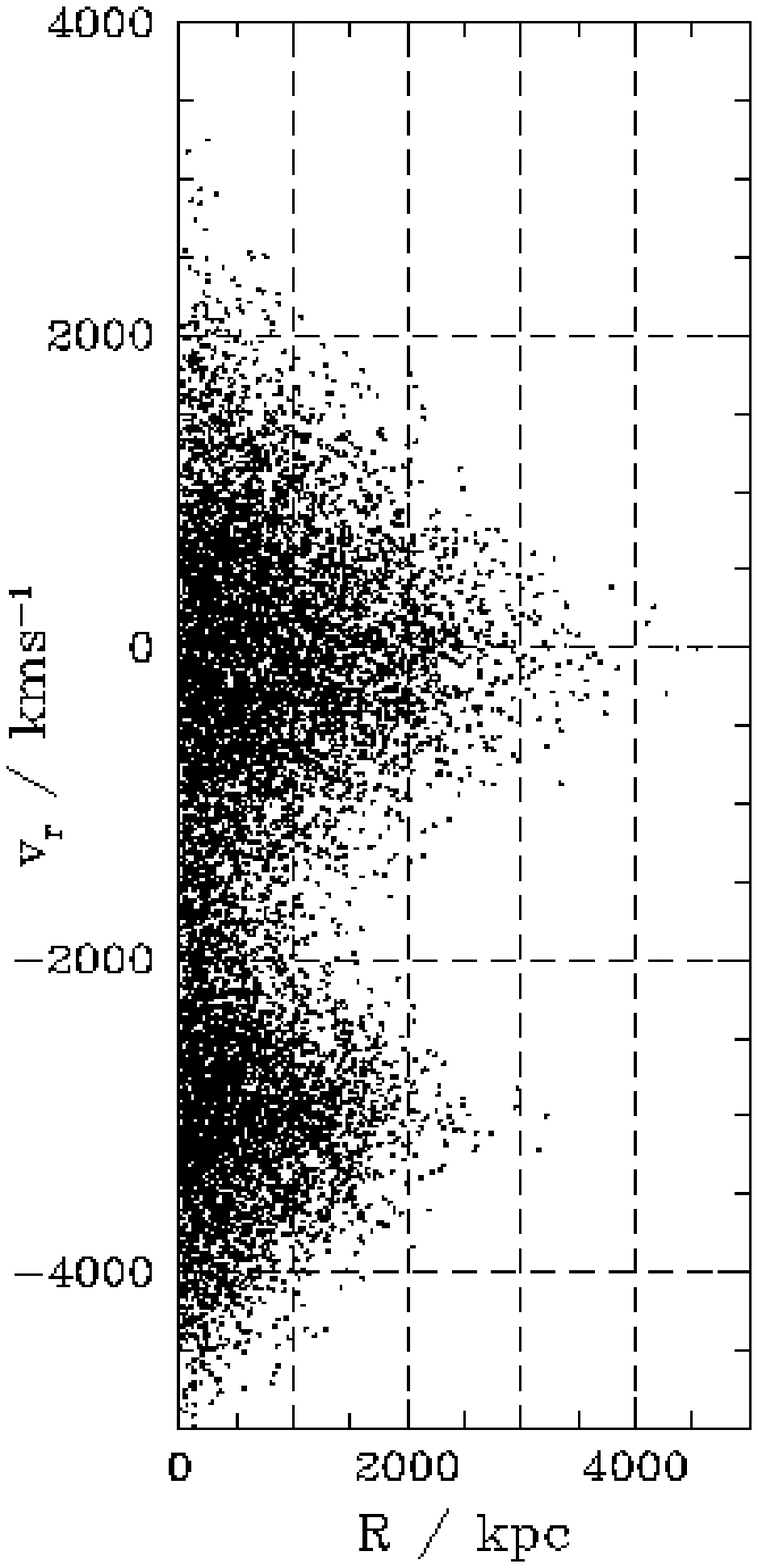}
    \includegraphics{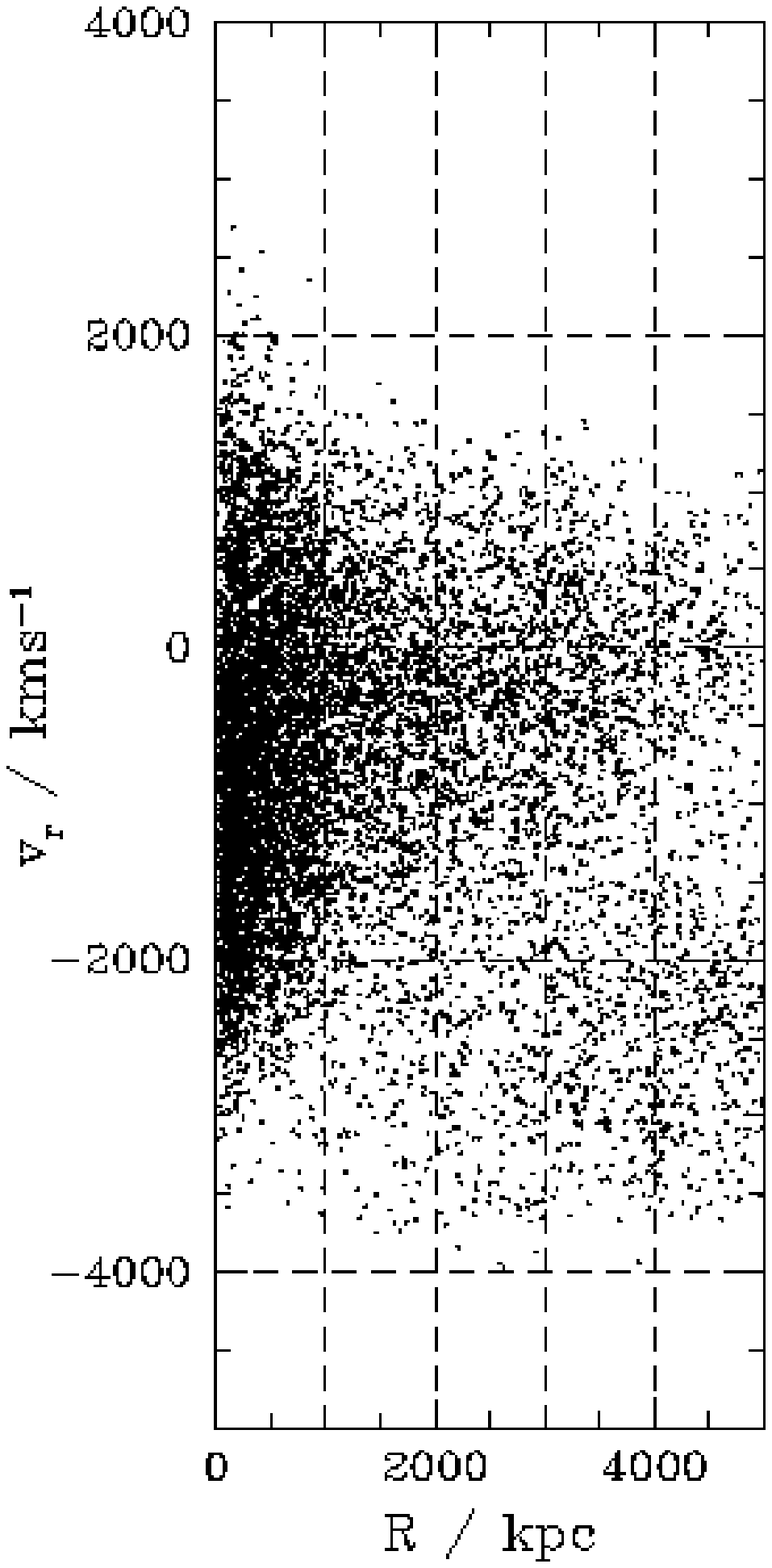}}
  \caption{The ``observed'' redshift distribution of particles in 
    the collision between the simulated clusters. The left panel shows the 
    initial conditions whilst the right panel shows the data 3 Gyrs after the 
    collision. This is to be compared with Fig.\ \ref{fig:scatterplot}.}
  \label{fig:redshift}
\end{figure}

The redshift distribution of $10\,000$ randomly selected particles at the final
time is shown in the right hand panel in Fig.\ \ref{fig:redshift}.
This is not a perfect match to the observational data but the main
features are present: we see a foreground component of ``galaxies''
that span large projected distances from the central region of
Cl0024+1654.  This component is separated by $\approx
3000\,\mathrm{km\,s}^{-1}$ in redshift space from the main
component. The smaller cluster bound core is moving away from the main
cluster at $\approx 1000\,\mathrm{km\,s}^{-1}$ -- the reduction in  
speed from the initial velocity is due to dynamical friction. A better
agreement would result if the impacting cluster was slightly more
massive in which case it would suffer more friction and would be seen
nearly at rest compared to the main component.  Also note that we are
plotting ``dark matter particles'' not ``galaxies''. Of course the
initial configuration shown in the left panel in Fig.\
\ref{fig:redshift} also displays a bimodal distribution of
redshifts. However, this configuration is more symmetric than either
the final simulated configuration or the observed redshift
distribution; also, the radial extent, in particular of the foreground
(in redshift space) component, is much smaller than observed. 

\begin{figure}
  \resizebox{\hsize}{!}{\includegraphics{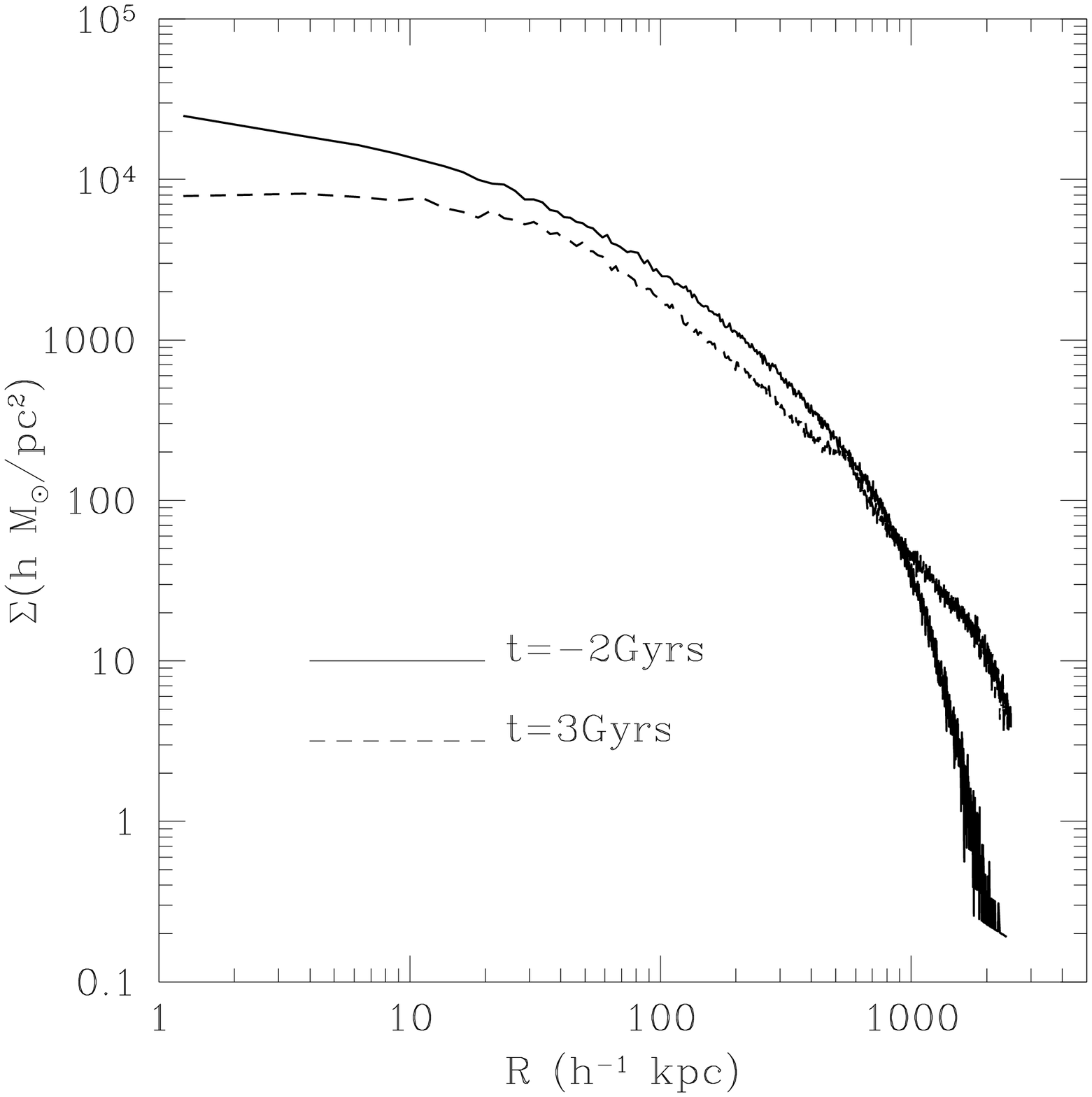}}
  \caption{The surface mass density profiles projected along the
    merger axis before (solid curve) and after the collision (dashed
    curve). The central surface mass density after the collision 
    is very close to that measured by \citet{Tyson1998}.
    }
  \label{fig:proj}
\end{figure}

The energy from the impulsive tidal shock has been transferred into
kinetic energy of the particles, heating and expanding the cold
central cores of the clusters. This leads to a flattening of the
physical and projected density profiles. In Fig.\ \ref{fig:proj}
we show the projected surface mass density profile before and after
the encounter. The initial profiles are cuspy, CDM type density
profiles whereas the final profiles have nearly constant density cores
in good agreement with that inferred from the mass reconstruction
\citep[their Figure 4]{Tyson1998}. The central surface mass density is
$4000\,M_\odot\,\mathrm{pc}^{-2}$ which is very close to the value
obtained by Tyson et al.\ for
$H_0=50\,\mathrm{km\,s}^{-1}\mathrm{Mpc}^{-1}$.  The 
total projected mass within the central $106\,\mathrm{kpc}$ is about
30\% lower than obtained by \citet{Broadhurst2000} which could be
reconciled by using a more massive encounter and/or properly including
the baryonic matter.  

\section{Comparison to other observations}
\label{sec:discussion}

The galaxy distribution of Cl0024+1654 in redshift space and on the
sky, as described in Sect.\ \ref{sec:environment}, provides strong
hints that this is not a simple relaxed system. How do the skewed
central velocity dispersion and the presence of a foreground component
in redshift space affect the interpretation of other observations and
in particular mass estimates of this system?

\subsection{Galaxy dynamics}
\label{ssec:dynamics}

Dynamical mass estimates rely on the assumption that the galaxies are
in virial equilibrium in the cluster's gravitational potential well
and that their velocities are purely random.

\begin{figure}
  \resizebox{\hsize}{!}{\includegraphics{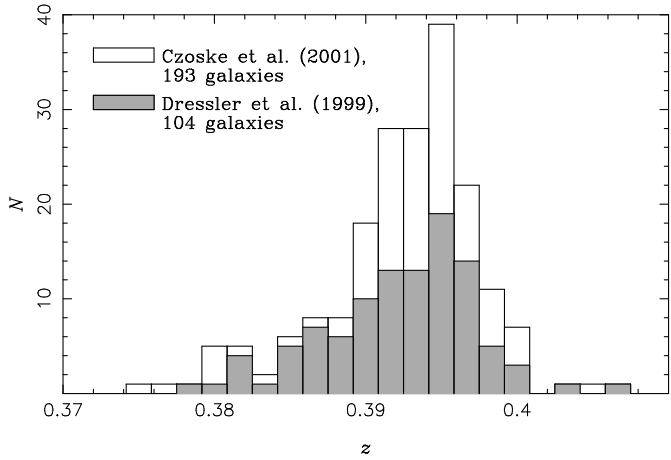}}
  \caption{Comparison of redshift histograms of the data of D99
    (corrected for the misidentifications discussed in Paper I) and
    Paper I (these include the data of D99), constrained to the area
    covered by D99, clustercentric distance $<\!5\arcmin$. Bin widths
    and centres are the same as in the inset of Figure 2 of D99.} 
  \label{fig:D99hist}   
\end{figure}

The large velocity dispersion of $\sigma \approx
1200\,\mathrm{km\,s}^{-1}$ found by \citet{Dressler-Gunn1992} and
\citet{Dressler1999} (hereafter D99) was based on galaxies within
$5\arcmin$ from the projected cluster centre. As shown in Fig.\
\ref{fig:D99hist}, there is no direct evidence for the bimodality of
Cl0024+1654 when attention is restricted to the central regions of the
cluster. This is still true with the larger number of galaxy spectra
used here: Whereas our catalogue contains about 85\% more redshifts in
this area than D99, the general shape of the histogram is the
same. Both histograms are strongly skewed towards low redshifts, thus
indicating the presence of substructure \citep{Ashman1994} or a bulk
velocity component. Note that the bimodality of the redshift
distribution only becomes apparent when galaxies at larger projected
distances from the cluster centre are included. In this case it is
therefore not the number of galaxy redshifts but their distribution
over a wide field that provides the clues to the complexity of this
system.   

The formal velocity dispersion for the 193 galaxies within $5\arcmin$
(Fig.\ \ref{fig:D99hist})is $\sigma_{\rm
  cent}\!=\!1050\,\mathrm{km\,s}^{-1}$. If there is indeed a bulk 
velocity component present in the central velocity distribution, then
mass estimates based on the formal central velocity dispersion will
overestimate the true cluster mass. 

A better dynamical mass estimate might be obtained from the velocity
dispersion at larger radii, where the two components A and B are
clearly separated. At projected distances $>\!3\arcmin$ the velocity
distribution of component A is regular with a dispersion 
of $\sim 600\,\mathrm{km\,s}^{-1}$. If we interpreted this value as the
velocity dispersion of a relaxed cluster, it would have only about a
quarter of the mass previously estimated for Cl0024+1654
\citep{Schneider1986}, below mass estimates from gravitational lensing
(which are roughly consistent with a high velocity dispersion) and even
below mass estimates from X-ray observations. 
In the context of the collision scenario presented in Section
\ref{sec:scenario} on the other hand, the galaxies at large projected
distance from the cluster centre are also affected by the collision
and cannot be used to derive a mass estimate based on the assumption
of dynamical equilibrium.

\subsection{X-ray observations}
\label{ssec:X-ray}

The X-ray observations from \textsc{Rosat}/HRI and ASCA
\citep{Soucail2000, Boehringer2000} by themselves show no indication
of anything other than a fairly small cluster of galaxies. Given the
measured values for Cl0024+1654,    
$T_\mathrm{X} = 5.7^{+4.9}_{-2.1}\,\mathrm{keV}$ and 
$L_\mathrm{X} = 6.8\times10^{43}\,h^{-2}\,\mathrm{erg\,s}^{-1}$
\citep{Soucail2000}, the gas does not seem too far away from 
the $L_\mathrm{X}-T_\mathrm{X}$ relation of \citet{Markevitch1998}; it
is slightly too hot for its luminosity. The morphology is regular and
the surface brightness profile well fitted by a beta profile with a
surprisingly small core radius of $33\,h^{-1}\,\mathrm{kpc}$. Whereas
no significant substructure is required to model the X-ray
observations of Cl0024+1654, they do not provide evidence against the
collision scenario either. 

The X-ray emission is difficult to predict in the context of two
clusters projected along the line-of-sight several Gyr after a head-on
collision. During the collision, hydrostatic equilibrium in the gas
component of the cluster(s) breaks down and the X-ray luminosity and
emission-weighted temperature fluctuate considerably (by up to a
factor of 10 for $L_{\rm X}$) and rapidly as a consequence of the
formation of shock waves and repeated expansion and compression of the
core gas \citep{Takizawa1999, Ricker-Sarazin2001, Ritchie-Thomas2001}.
After several Gyr however, shocks have dissipated and the gas settles
down to an equilibrium configuration. The cited hydrodynamic
simulations of cluster mergers consider collision speeds of about
$1000\,\mathrm{km\,s}^{-1}$; in this case the dark matter cores
separate to distances of only a couple of Mpc before turning around
and eventually merging into one clump. In a high-speed collision,
however, the collisionless dark matter cores of the two clusters
separate to large distance before turning around. The gas, due to its
collisional nature should experience stronger interaction during the
first crossing and thus behave quite differently from the dark
matter/galaxies and possibly also from what low-speed collision
simulations predict. \citet{Ricker-Sarazin2001} observe a slight
segregation between the gas and the dark matter cores in their
simulations. This effect should be more pronounced in a high-speed
collision.

If we accept the collision scenario then a mass estimate from the
X-ray observations is not possible. Detailed inclusion of a
hydrodynamic treatment of the gas component in simulations of
high-speed collisions is necessary to understand the behaviour of the
X-ray emission in this case. High resolution X-ray imaging with
\textsc{Chandra} has recently been obtained (P.~I.\ Hattori) but has
not been published yet.

\subsection{Gravitational lensing}
\label{ssec:gravitational lensing}

Gravitational lensing by clusters of galaxies does not rely on the
matter in the clusters being in dynamical equilibrium. However,
gravitational lensing measures the weighted integral of all the mass
between the observer and the source, and the interpretation of the
measured mass value thus depends on the detailed distribution of mass
along the line-of-sight.The skew in the central redshift distribution
in Cl0024+1654 is an indication of bulk motion, possibly due to a
high-speed collision between two fairly massive clusters of a mass
ratio of roughly 2:1, or by the passage of a small group through the
core of Cl0024+1654.  In the latter case the mass determined by strong
lensing models would indeed be indicative of the mass of the main
cluster; however, in this case one would expect the X-ray emitting gas
to be largely unperturbed by the collision, and one would be left with
the discrepancy between the mass estimates derived from lensing and
X-rays. In the former case, the lensing mass would be the sum of the
two cluster cores and therefore too large if interpreted as
representing the mass of a single cluster.

The projected mass profile for the post-merger system seen in our
simulations shows two characteristics which are testable by
gravitational lensing analyses of Cl0024+1654: At small scales of
several 10's kpc, the profile shows a flat core which is consistent
with strong lensing mass models \citep{Tyson1998}, as shown in Sect.\
\ref{sec:scenario}. On large scales the profile is significantly
flatter than the initial CDM profile, falling only as $r^{-2.5}$ out
to $\sim\!3\,\mathrm{Mpc}$. Previous weak lensing analyses on
Cl0024+1654 have not attempted a detailed reconstruction of the mass
profile. \citet{Bonnet1994} and \citet{vanWaerbeke1997} derived shear
maps from ground-based data on the north-east quadrant of the cluster,
\citet{Smail1997} were restricted to the small field of view of
HST/WFPC2. The shear pattern found by \citet{Bonnet1994} is compatible
with an isothermal sphere profile, $\rho\propto r^{-2}$ out to
$3\,h^{-1}\,\mathrm{Mpc}$. 

A weak lensing analysis of newly obtained CFHT/CFH12k images in BVRI
is currently underway (Czoske et al.\ 2002, in preparation). Also, a
sparsely sampled HST/WFPC2 mosaic consisting of 38 pointings out to a
distance of $2.5\,h^{-1}\,\mathrm{Mpc}$ from the cluster centre has
recently been obtained; these data will also be used for a weak
lensing analysis \citep{Treu2001}. These data will allow to probe the
cluster's projected radial mass profile accurately to the edge of the
survey field.

\section{Discussion and conclusions}
\label{sec:conclusions}

In the present paper we have analyzed the galaxy distribution in the
cluster of galaxies Cl0024+1654, based on $\sim\!300$ galaxy redshifts
and projected positions. The cluster, which was previously regarded as
a prototype of a massive relaxed cluster at intermediate redshift,
turns out to have a fairly complicated structure, showing a strongly
skewed redshift distribution in its central parts and two
well-separated components at larger projected distances out to
$\sim\!3\,h^{-1}\mathrm{Mpc}$.

We argue that the blue tail of the central velocity distribution and
the foreground component originate from the same physical system and
interpret the peculiar redshift-space distribution as the result of a
high-speed head-on collision of two clusters of galaxies, the merger
axis being very nearly parallel to the line-of-sight. Using a
numerical simulation we show that it is possible to explain the
observed redshift distribution with a high-speed collision of two
rather massive clusters of galaxies with a mass ratio of about 2:1.

Apart from reproducing the spatial/redshift distribution of the
cluster galaxies, this scenario also produces a projected mass
distribution which is very close to that derived by \citet{Tyson1998}
from the quintuple arc system observed in Cl0024+1654. Note that the
scenario was not designed to reproduce this mass profile. Since it is
thus possible to produce a mass distribution with a flat core from the
merger of two CDM halos this eliminates one of the main arguments
against simple non-interacting cold dark matter as the dynamically
dominant component in clusters of galaxies
\citep{Spergel-Steinhardt2000}.

The distribution of spectral types in the foreground component B is
more akin to the general field population than to a cluster
population. In the context of our collision scenario, these would
correspond to the outer regions of the smaller cluster which have
become unbound during the impact. Even initially, these galaxies would
probably not correspond to a fully transformed cluster galaxy
population. In addition, the impact might have triggered additional
star formation in these galaxies. Remembering that Cl0024+1654 was one
of the most distant clusters in which the Butcher-Oemler effect was
detected \citep{Butcher-Oemler1984, Dressler-Gunn1985}, this may add a
new view on this effect: The fact that $\sim40$\% of the bright
galaxies in the cluster are emission line galaxies
([\ion{O}{ii}]$>5\,$\AA) has a natural interpretation in the collision
scenario. The spectral distribution of the galaxies in component A
(and centre) are also perturbed by the collision with a possible
excess of rHDS and SF/SBB (Tables \ref{tab:oii-hd} and
\ref{tab:br-hd}); rHDS galaxies may represent the result of a
starburst induced during the early stages of the interaction.

The relative velocity of the clusters of
$\sim\!3000\,\mathrm{km\,s}^{-1}$ implied by the redshift distribution
in Cl0024+1654 is very high. Observations and simulations find mean
peculiar velocities for clusters of order $500\,\mathrm{km\,s}^{-1}$
\citep{Bahcall-Oh1996, Giovanelli1998, Gibbons2001, Colberg2000a}.
Still, colliding clusters can reach relative velocities of about
$3000\,\mathrm{km\,s}^{-1}$ at separations of about $1\,\mathrm{Mpc}$
\citep{Sarazin2001}. \citet{Markevitch2001} have recently found a bow
shock in the galaxy cluster 1E0657-56 which implies a relative speed
of 3000 to $4000\,\mathrm{km\,s}^{-1}$ for the collision. Cl0024+1654
is still an extraordinary but not an impossible system.

\begin{acknowledgements}
  OC thanks the European Commission for generous support under grant
  number ER-BFM-BI-CT97-2471 and the Institute for Astronomy in
  Honolulu, in particular Harald Ebeling, for kind hospitality. JPK
  thanks CNRS for support. We also acknowledge support from the
  UK-French ALLIANCE collaboration programme 0161XM and support from
  the TMR Network ``Gravitational Lensing: New Constraints on
  Cosmology and the Distribution of Dark Matter'' of the European
  Commission under contract No.\ ER-BFM-RX-CT97-0172. We thank the
  referee, Richard Ellis, for his comments which helped improve the
  paper.
\end{acknowledgements}

\bibliography{Cl0024,articles,books,unpublished}

\begin{thebibliography}{47}
\expandafter\ifx\csname natexlab\endcsname\relax\def\natexlab#1{#1}\fi

\bibitem[{Ashman {et~al.}(1994)Ashman, Bird, \& Zepf}]{Ashman1994}
Ashman, K.~A., Bird, C.~M., \& Zepf, S.~E. 1994, \aj, 108, 2348

\bibitem[{Bahcall \& Oh(1996)}]{Bahcall-Oh1996}
Bahcall, N.~A. \& Oh, S.~P. 1996, \apj, 462, L49

\bibitem[{Balogh {et~al.}(1999)Balogh, Morris, Yee, Carlberg, \&
  Ellingson}]{Balogh1999}
Balogh, M.~L., Morris, S.~L., Yee, H. K.~C., Carlberg, R.~G., \& Ellingson, E.
  1999, \apj, 527, 54

\bibitem[{Beers {et~al.}(1990)Beers, Flynn, \& Gebhardt}]{Beers1990}
Beers, T.~C., Flynn, K., \& Gebhardt, K. 1990, \aj, 100, 32

\bibitem[{B{\" o}hringer {et~al.}(2000)B{\" o}hringer, Soucail, Mellier, Ikebe,
  \& Schuecker}]{Boehringer2000}
B{\" o}hringer, H., Soucail, G., Mellier, Y., Ikebe, Y., \& Schuecker, P. 2000,
  \aap, 353, 124

\bibitem[{Bonnet {et~al.}(1994)Bonnet, Mellier, \& Fort}]{Bonnet1994}
Bonnet, H., Mellier, Y., \& Fort, B. 1994, \apj, 427, L83

\bibitem[{Broadhurst {et~al.}(2000)Broadhurst, Huang, Frye, \&
  Ellis}]{Broadhurst2000}
Broadhurst, T., Huang, X., Frye, B., \& Ellis, R.~S. 2000, \apj, 534, L15

\bibitem[{Butcher \& Oemler(1984)}]{Butcher-Oemler1984}
Butcher, H. \& Oemler, Jr., A. 1984, \apj, 285, 426

\bibitem[{Colberg {et~al.}(2000)Colberg, White, MacFarland, Jenkins, Pearce,
  Frenk, Thomas, \& Couchman}]{Colberg2000a}
Colberg, J.~M., White, S. D.~M., MacFarland, T.~J., {et~al.} 2000, \mnras, 313,
  229

\bibitem[{Czoske {et~al.}(2001)Czoske, Kneib, Soucail, Bridges, Mellier, \&
  Cuillandre}]{Czoske2001a}
Czoske, O., Kneib, J.-P., Soucail, G., {et~al.} 2001, \aap, 372, 391

\bibitem[{Dressler \& Gunn(1992)}]{Dressler-Gunn1992}
Dressler, A. \& Gunn, J.~E. 1992, \apjs, 78, 1

\bibitem[{Dressler {et~al.}(1985)Dressler, Gunn, \&
  Schneider}]{Dressler-Gunn1985}
Dressler, A., Gunn, J.~E., \& Schneider, D.~P. 1985, \apj, 294, 70

\bibitem[{Dressler {et~al.}(1999)Dressler, Smail, Poggianti, Butcher, Couch,
  Ellis, \& Oemler}]{Dressler1999}
Dressler, A., Smail, I., Poggianti, B.~M., {et~al.} 1999, \apjs, 122, 51

\bibitem[{Ghigna {et~al.}(2000)Ghigna, Moore, Governato, Lake, Quinn, \&
  Stadel}]{Ghigna2000}
Ghigna, S., Moore, B., Governato, F., {et~al.} 2000, \apj, 544, 616

\bibitem[{Gibbons {et~al.}(2001)Gibbons, Fruchter, \& Bothun}]{Gibbons2001}
Gibbons, R.~A., Fruchter, A.~S., \& Bothun, G.~D. 2001, \aj, 121, 649

\bibitem[{Giovanelli {et~al.}(1998)Giovanelli, Haynes, Salzer, Wegner,
  da~Costa, \& Freudling}]{Giovanelli1998}
Giovanelli, R., Haynes, M.~P., Salzer, J.~J., {et~al.} 1998, \aj, 116, 2632

\bibitem[{Girardi \& Mezzetti(2001)}]{Girardi-Mezzetti2001}
Girardi, M. \& Mezzetti, M. 2001, \apj, 548, 79

\bibitem[{Hogan \& Dalcanton(2000)}]{Hogan-Dalcanton2000}
Hogan, C.~J. \& Dalcanton, J.~J. 2000, \prd, 62, 063511

\bibitem[{Kassiola {et~al.}(1992)Kassiola, Kovner, \& Fort}]{Kassiola1992}
Kassiola, A., Kovner, I., \& Fort, B. 1992, \apj, 400, 41

\bibitem[{Markevitch(1998)}]{Markevitch1998}
Markevitch, M. 1998, \apj, 504, 27

\bibitem[{Markevitch {et~al.}(2001)Markevitch, Gonzalez, David, Vikhlinin,
  Murray, Forman, Jones, \& Tucker}]{Markevitch2001}
Markevitch, M., Gonzalez, A.~H., David, L., {et~al.} 2001, astro-ph/0110468

\bibitem[{Markevitch {et~al.}(2000)Markevitch, Ponman, Nulsen, Bautz, Burke,
  David, Davis, Donnelly, Forman, Jones, Kaastra, Kellogg, Kim, Kolodziejczak,
  Mazzotta, Pagliaro, Patel, Van~Speybroeck, Vikhlinin, Vrtilek, Wise, \&
  Zhao}]{Markevitch2000}
Markevitch, M., Ponman, T., Nulsen, P. E.~J., {et~al.} 2000, \apj, 541, 542

\bibitem[{Markevitch \& Vikhlinin(2001)}]{Markevitch-Vikhlinin2001}
Markevitch, M. \& Vikhlinin, A. 2001, astro-ph/0105093

\bibitem[{Mazzotta {et~al.}(2001)Mazzotta, Markevitch, Vikhlinin, Forman,
  David, \& VanSpeybroek}]{Mazzotta2001a}
Mazzotta, P., Markevitch, M., Vikhlinin, A., {et~al.} 2001, \apj, 555, 205

\bibitem[{Miralda-Escud{\'e} \& Babul(1995)}]{Miralda-Escude-Babul1995}
Miralda-Escud{\'e}, J. \& Babul, A. 1995, \apj, 449, 18

\bibitem[{Moore {et~al.}(2000)Moore, Gelato, Jenkins, Pearce, \&
  Quilis}]{Moore2000}
Moore, B., Gelato, S., Jenkins, A., Pearce, F.~R., \& Quilis, V. 2000, \apj,
  535, L21

\bibitem[{Moore {et~al.}(1998)Moore, Governato, Quinn, Stadel, \&
  Lake}]{Moore1998}
Moore, B., Governato, F., Quinn, T., Stadel, J., \& Lake, G. 1998, \apj, 499,
  L5

\bibitem[{Navarro {et~al.}(1997)Navarro, Frenk, \& White}]{Navarro1997}
Navarro, J.~F., Frenk, C.~S., \& White, S. D.~M. 1997, \apj, 490, 493

\bibitem[{Press {et~al.}(1992)Press, Teukolsky, Vetterling, \&
  Flannery}]{Press1992}
Press, W.~H., Teukolsky, S.~A., Vetterling, W.~T., \& Flannery, B.~P. 1992,
  Numerical Recipes in {C}, 2nd edn. (Cambridge University Press)

\bibitem[{Ricker \& Sarazin(2001)}]{Ricker-Sarazin2001}
Ricker, P.~M. \& Sarazin, C.~L. 2001, astro-ph/0107210

\bibitem[{Ritchie \& Thomas(2001)}]{Ritchie-Thomas2001}
Ritchie, B.~W. \& Thomas, P.~A. 2001, astro-ph/0107374

\bibitem[{Sarazin(2001)}]{Sarazin2001}
Sarazin, C.~L. 2001, in Merging Processes in Clusters of Galaxies, ed.
  L.~Feretti, I.~M. Gioia, \& G.~Giovannini (Kluwer, Dordrecht),
  astro-ph/0105418

\bibitem[{Schneider {et~al.}(1986)Schneider, Dressler, \& Gunn}]{Schneider1986}
Schneider, D.~P., Dressler, A., \& Gunn, J.~E. 1986, \aj, 92, 523

\bibitem[{Shapiro \& Iliev(2000)}]{Shapiro-Iliev2000}
Shapiro, P.~R. \& Iliev, I.~T. 2000, \apj, 542, L1

\bibitem[{Shapiro \& Wilk(1965)}]{Shapiro-Wilk1965}
Shapiro, S.~S. \& Wilk, M.~B. 1965, Biometrika, 52, 591

\bibitem[{Silverman(1986)}]{Silverman1986}
Silverman, B.~W. 1986, Density Estimation for Statistics and Data Analysis,
  Monographs on Statistics and Applied Probability (London: Chapman and Hall)

\bibitem[{Smail {et~al.}(1996)Smail, Dressler, Kneib, Ellis, Couch, Sharples,
  \& Oemler}]{Smail1996}
Smail, I., Dressler, A., Kneib, J.-P., {et~al.} 1996, \apj, 469, 508

\bibitem[{Smail {et~al.}(1997)Smail, Ellis, Dressler, Couch, Oemler, Sharples,
  \& Butcher}]{Smail1997}
Smail, I., Ellis, R.~S., Dressler, A., {et~al.} 1997, \apj, 479, 70

\bibitem[{Soucail {et~al.}(2000)Soucail, Ota, B{\"o}hringer, Czoske, Hattori,
  \& Mellier}]{Soucail2000}
Soucail, G., Ota, N., B{\"o}hringer, H., {et~al.} 2000, \aap, 355, 433

\bibitem[{Spergel \& Steinhardt(2000)}]{Spergel-Steinhardt2000}
Spergel, D.~N. \& Steinhardt, P.~J. 2000, \prl, 84, 3760

\bibitem[{Takizawa(1999)}]{Takizawa1999}
Takizawa, M. 1999, \apj, 520, 514

\bibitem[{Treu {et~al.}(2001)Treu, Ellis, Trivedi, Kneib, Dressler, Oemler,
  Natarajan, \& Smail}]{Treu2001}
Treu, T., Ellis, R.~S., Trivedi, P., {et~al.} 2001, in SESTO-2001-Tracing
  Cosmic Evolution with Galaxy Clusters, ASP conference series,
  astro-ph/0112409

\bibitem[{Tyson {et~al.}(1998)Tyson, Kochanski, \& {Dell'Antonio}}]{Tyson1998}
Tyson, J.~A., Kochanski, G.~P., \& {Dell'Antonio}, I.~P. 1998, \apj, 498, L107

\bibitem[{van Waerbeke {et~al.}(1997)van Waerbeke, Mellier, Schneider, Fort, \&
  Mathez}]{vanWaerbeke1997}
van Waerbeke, L., Mellier, Y., Schneider, P., Fort, B., \& Mathez, G. 1997,
  \aap, 317, 303

\bibitem[{Vikhlinin {et~al.}(2001)Vikhlinin, Markevitch, \&
  Murray}]{Vikhlinin2001}
Vikhlinin, A., Markevitch, M., \& Murray, S.~S. 2001, \apj, 551, 160

\bibitem[{Wu {et~al.}(1998)Wu, Chiueh, Fang, \& Xue}]{Wu1998}
Wu, X.-P., Chiueh, T., Fang, L.-Z., \& Xue, Y.-J. 1998, \mnras, 301, 861

\bibitem[{Yee {et~al.}(1996)Yee, Ellingson, \& Carlberg}]{Yee1996a}
Yee, H. K.~C., Ellingson, E., \& Carlberg, R.~G. 1996, \apj, 102, 269

\end{thebibliography}

\end{document}